\documentclass[11pt]{amsart}
\usepackage[hyphens]{url}
\usepackage[english]{babel}
\usepackage{amsmath,amsthm,caption}
\usepackage{amsfonts, natbib,graphicx}

\theoremstyle{definition}

\theoremstyle{remark}

\begin{document}

\title[QKD Overview]{An Overview of the State of the Art for Practical Quantum Key Distribution}

\author[Moskovich]{Daniel D. Moskovich}
\date{9 April, 2015}

\address{Center for Quantum Information Science and Technology, Ben-Gurion University of the Negev, Israel}%



\begin{abstract}
This is an overview of the state of the art for quantum key distribution (QKD) as of March 2015. It is written by a non-expert for non-experts. Additions and corrections are welcome.
\end{abstract}

\maketitle

\section{Introduction}


The goal of this overview to concisely summarize, in a way that is accessible to a non-expert, where practical Quantum Key Distribution (QKD) stands now in early 2015 and what seem to be promising directions for the near future to the best of the author's knowledge and understanding.

We begin with a general overview of what QKD is, followed by a discussion of the major practical QKD players at the moment, a discussion of protocols, and a discussion of photon sources, transmission, and detection. We conclude with a section on attacks against QKD.

\section{What is quantum key distribution?}

Quantum key distribution (QKD) uses principles of quantum information theory to ensure secure communication \citep{Weisner:83,BB:84,E:91}. Its goal is for two parties called (A)lice and (B)ob to share a secret key made up of $0$\,s and $1$\,s which they will later use to encrypt and decrypt communications between them. The information used to compose the key is carried between Alice and Bob on qubits (two state quantum systems). 

The \emph{SECOCQ White Paper} convincingly argued that QKD is a form of \emph{trusted courier} (Alice hands a message to somebody she trusts, who carries that to Bob), so that it is useful in contexts in which a trusted courier would be useful \citep{SECOQC:07}. With idealized hardware and with perfect accuracy, the advantages of QKD over classical trusted courier methods would be: 

\begin{itemize}
\item Mathematically-proven security against all classical and quantum attacks.
\item Alice and Bob can detect any active attempt by an eavesdropper (E)ve to eavesdrop on the key distribution process.
\end{itemize}

The \emph{Black Paper of Quantum Cryptography} convincingly argued that real-life QKD security is less than perfect, so that each different QKD setup should be carefully and individually studied to assure its security \citep{ScaraniKurtsiefer:14}. All security threats discovered so far have been redeemable, and we have no reason to believe that any given QKD setup cannot be made perfectly secure against all known attacks in principle.

Real world QKD has become a focus of interest for industrial players, for governments, and for security agencies.

\section{Fundamental challenges}

A number of fundamental challenges to the widescale use of QKD have been identified \citep{PritchardTill:10}.

\begin{enumerate}
\item Limited transmission rate and range. Both the range and the maximal bit-rate of QKD are low compared to classical communications. It is considered technologically impossible, for instance, to transmit a polarized photon reliably over more than 400km of fiber, although quantum repeaters will allow for longer range QKD.

\item QKD protocols are fundamentally point-to-point, and do not integrate with packet-based protocols such as those used on the internet.

\item QKD requires expensive special-purpose hardware such as single-photon sources and detectors. Such hardware is difficult to upgrade and to maintain.

\item QKD addresses only one aspect of the security problem. For example authentication and integrity are not covered and must be handled classically.

\item There is nothing fundamentally wrong with existing classical cryptographic techniques. Even if some classical ciphers (\textit{e.g.} RSA) may be cracked using quantum algorithms at some unspecified point in the future, other classical ciphers are being developed that would be immune to quantum attacks.

\item Because it is a new technology, there are potential discovered and undiscovered vulnerabilities in practical QKD systems. Indeed, several proposed conditions for unconditional security of practical implementations were found wanting and had to be revised, \textit{e.g.} because key-length is finite while many security proofs assume infinite-length keys \citep{Inamori:07,Tomamichel:12}, or because the quantum effect of \emph{locking} might allow unexpectedly large information leak during the error-correction and privacy amplification steps \citep{Konig:07,Iwakoshi:14,Yuen:13,Portmann:14}. Commercial QKD systems have been successfully hacked (Section~\ref{SS:Trojan}). Because QKD is unconditionally secure while the security of any quantum protocol implementation is a probabilistic quantitative matter, these vulnerabilities will in principle never be fatal flaws. But each new vulnerability might require re-tuning of parameters and modifications to technological implementation.
\end{enumerate}

 \section{Advantages of QKD}

 \begin{enumerate}
 \item QKD provides the possibility to establish a secret key in a way that is provable secure against eavesdropping. Moreover, QKD can be composed with other encryptions, so as to provide an additional layer of security for an already secure message. For example a message encrypted using an RSA public key may be once again encrypted using a quantum key. To intercept the message, an attacker would have to break both the quantum key and the classical key.

 \item Eavesdropping can be detected, following which countermeasures may be adopted. This capability distinguished QKD from among all encryption methods.

 \item Expertise and knowledge gained in QKD research will, in large part, be useful for developing future technologies in future manifestations of the coming quantum revolution predicted by Michael Berry when he said, ``It is easy to predict that in the twenty-first century, it will be quantum mechanics that influences all our lives.'' \cite{Berry:98}.
 \end{enumerate}

\section{World QKD projects}


\subsection{Large scale networks}
\begin{figure}
\centering
\includegraphics[width=0.5\textwidth]{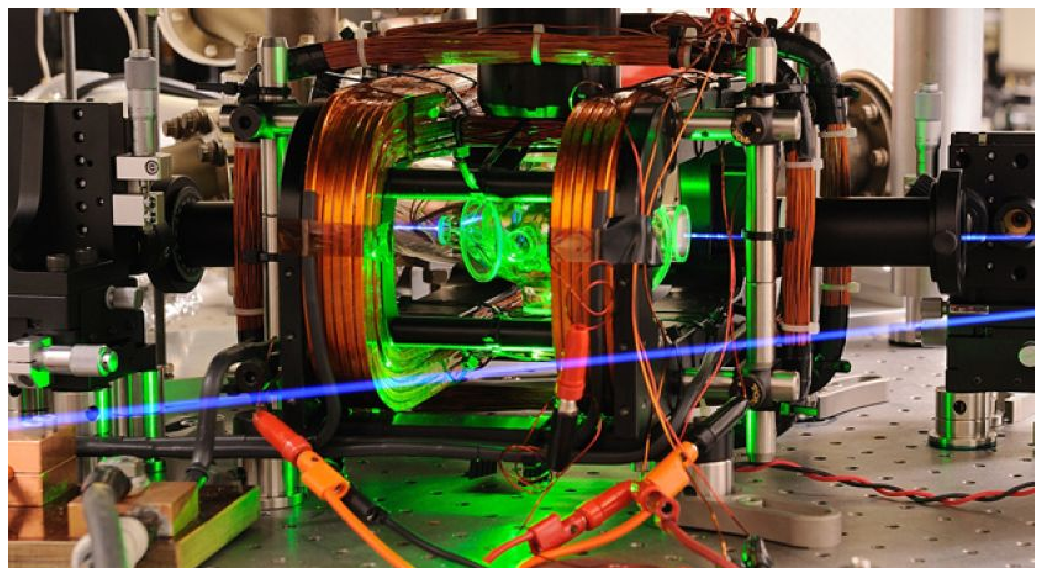}
\caption*{\small Microwave apparatus used in quantum experiments. Retrieved from \protect\url{http://www.foxnews.com/tech/2013/05/08/quantum-network-secretly-running-for-2-years/}}
\end{figure}

\begin{sloppypar}
In the last 10 years, a number of multi-user QKD networks have been constructed. All use relay between trusted nodes and optical switching. The first of these was the $10$--node DARPA Quantum Network which has been operating since 2004 \citep{Elliot:05}. It uses active optical switching (\textit{i.e.} an electrically powered switching device similar to a router) to distribute the key between pairs of nodes. It is being developed by BBN Technologies, Harvard University, Boston University and QinetiQ, with the support of the US Defense Advanced Research Projects Agency (DARPA).  The SECOQC (Secure Communication Based on Quantum Cryptography) Quantum Network is an EU project which integrated several different QKD systems into one quantum backbone (QBB) network, developing a cross-platform interface (\url{http://www.secoqc.net/}). This provided impetus for the European Telecommunications Standards Institute (ETSI) to launch an industry specification group for QKD (ISG-QKD)in order to create universally accepted QKD standards \citep{ETSI:15}. The Swiss Quantum Network and the Durban Network are testing long-term QKD operation in field environments (\url{http://swissquantum.idquantique.com} and \citep{Mirza:10}). Transparent network implementation of QKD using only beam splitters, which facilitate secure communication without requiring clients to be reconfigured, have been demonstrated by several groups (Telecordia Technologies, Universidad Polit\'{e}cnica de Madrid and Telef\'{o}nica Investigaci\'{o}n y Desarrollo, and two teams from the University of Science and Technology of China). The Tokyo QKD network used a central Key Management Service (KMS) and newer technologies to increase its speed to the point of transmitting a QKD-secured live teleconference between two nodes \citep{Sasaki:11}. This is suitable for government or municipal settings in which one central body controls the flow of information. Mitsubishi combined this system with an application for secure telephony to demonstrate QKD-secured mobile telephony \citep{Mitsubishi:15}. Finally, Los Alamos National Laboratory (LANL) runs a hub-and-spokes one-to-many quantum network \citep{Hughes:13}. The LANL photon generator has been miniaturized to around the size of a house key.
\end{sloppypar}

\begin{sloppypar}
China is currently constructing a 1200-mile line between Beijing and Shanghai as part of a proposed $20$-node QKD network which it aims to complete in 2016. Its current network, the Hefei-Chaohu-Wuhu wide area QKD network, is the largest in the world \citep{Wang:14}. To overcome the need to use trusted nodes, where one compromised node could impact the security of the entire network, there has been work aimed at using techniques of classical multiple access optical communication in the quantum context \citep{PashaRam:04}. Such technologies have been applied for one part of the DARPA network, and also for an experimental three-node network at NIST (see \url{http://www.nist.gov/itl/quantum/threeusernetwork.cfm}). 
\end{sloppypar}

Taking the above technologies into account, the Engineering Science and Research Council (EPSRC) estimated in their 2014 report that hand-held QKD systems should be commercially available ``with sufficient investment and encouragement'' within 4--7 years, and that long-range highly-connected quantum networks should become available within 10--25 years \citep{EPSRC:14}.

\subsection{University Centres}

There are a growing number of university centers in the world which specialize in quantum communication. We list a few of the most active.


\begin{sloppypar}
The world's foremost dedicated quantum communications center is the Group of Applied Physics (GAP) at Geneva University (\url{http://www.unige.ch/gap/quantum/}) and their commercial spinoff company Id Quantique (\url{http://www.idquantique.com/}). They have developed what is today the world's best single photon detector \citep{Korzh:14} with which they have achieved the current world record distance for QKD through fiber \citep{Korzh:15}. They also produce and sell photon detectors and random number generators using patented technologies.
\end{sloppypar}

\begin{sloppypar}
The Centre for Quantum Technologies (CQT) in Singapore, directed by Ekert who developed the E91 protocol, specializes in quantum hacking \url{http://www.quantumlah.org/}. They have developed several successful attacks, which have taken advantage both of side-channels (\textit{e.g.} \citep{LamasKurtsiefer:07}) and of erroneous parameters in security proofs (\textit{e.g.} \citep{Gerhardt:11}).
\end{sloppypar}

The Institute of Quantum Computing (IQC) in Waterloo also has a research group working on QKD. It is directed by Norbert L\"{u}tkenhaus, who previously worked at MagiQ to develop practical QKD. Vadim Makarov of that group discovered some successful side-channel attacks against QKD (\textit{e.g.} \citep{Makarov:06,Makarov:09}).

\begin{sloppypar}
The Key Laboratory of Quantum Information is China's leading quantum information center, which is creating the world's longest and most sophisticated QKD networks \url{http://en.physics.ustc.edu.cn/research_9/Quantum/201107/t20110728_116550.html}.
\end{sloppypar}

\begin{figure}
\centering
\includegraphics[width=0.5\textwidth]{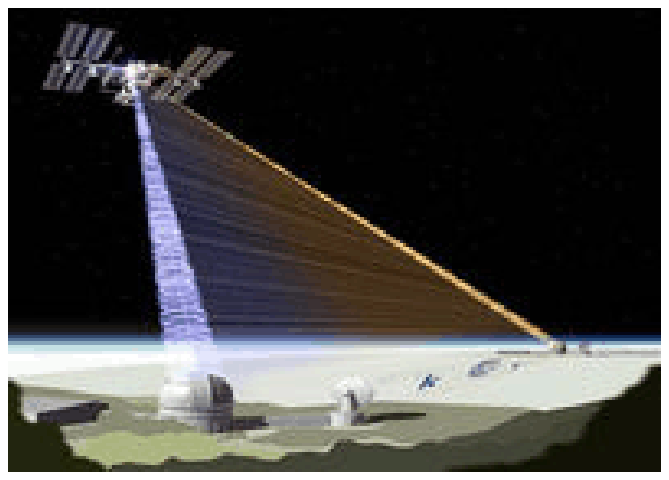}
\caption*{\small Artist's conception of quantum key distribution over free space ground to satellite quantum information links. Retrieved from \protect\url{http://www.esa.int/ESA}}
\end{figure}

\subsection{Commercial companies}

\begin{sloppypar}
A number of commercial companies sell QKD systems and related devices. MagiQ Technologies in the US sells the QPN-8505, a QKD system which combines BB84 QKD with classical 3DES and AES encryption (\url{http://www.magiqtech.com/}). It works using decoy-state optimized BB84, with a secure key rate of 256Hz over 100km, or 140km with decoy states. In Europe the leading QKD company is Id Quantique, whose flagship product is the Clavis2, a pure QKD system (\url{http://www.idquantique.com/}). Clavis2 implements both BB84 and SARG04, with secure key-rates of around 1KHz on a 25km fiber. SeQureNet is a Paris-based company that produces QKD parts and that specialized is continuous variable (CV) QKD (\url{http://sequrenet.com}). Quintessence Labs in Australia provides true random number generators (\url{http://www.quintessencelabs.com/}).
\end{sloppypar}

\section{Protocols}

\subsection{BB84}

The most widely used QKD protocol, which was also the world's first QKD protocol, was developed by Bennett and Brassard in 1984, and is called BB84.  It is typically divided into three layers: The physical layer in which the quantum communication is carried out,the key-extraction in which the actual key is extracted from the qubits that Alice sent to Bob, and the key-application layer where the secret key is used to encode a communication such as a telephone or a video conversation \citep{BB:92,Gisin:02}.

In the physical layer, Alice sends random photons, $1$ with 50\% probability and $0$ with 50\% probability, either in the so-called $X$ basis or in the so-called $Z$ basis, each with 50\% probability. Bob measures each bit he receives in a random basis, either in the $X$ basis with 50\% probability, or in the $Y$ basis with 50\% probability. This is the hardware-intensive portion of the protocol, for which good random-number generators, single-photon sources, and single-photon detectors are required.

In the key-extraction layer, BB84 becomes classical. The first classical sublevel is called \emph{sifting}. Alice and Bob both reveal which bases they used over a public channel. They then discard the bits which they measured in different bases. The second sublevel is called \emph{authentication}. In it, Alice and Bob compare some of their sifted bits over the public channel to determine whether eavesdropping has occurred. If the bits they compare are more different than can be accounted by from random noise, then they can guess that Eve has eavesdropped, and adopt countermeasures. The reason that they can make this deduction is that Eve's direct attack, intercept--resend REF, would involve measuring some of the bits sent by Alice, and sending them on to Bob. But since Eve does not know which basis was originally used by Alice, she will choose the wrong basis with 50\% probability, and if she chooses the wrong basis then she will send the wrong qubit to Bob, and that incorrect qubit will survive into the sifted key with 50\% probability. The third sublevel is called \emph{error correction}. In it, Alice and Bob apply classical error-correcting algorithms to remedy the effect of random errors caused by channel noise and by the fact that equipment is non-ideal. The fourth and final level is called \emph{privacy amplification}, in which Alice and Bob apply classical cryptography algorithms to minimize the effect on the final key of any under-the-radar eavesdropping by Eve. In other words, security of a QKD key is always a quantitative affair because of non-ideal equipment and channel noise, so some non-trivial information might have been picked-up by Eve without being detected in the authentication phase. But the amount of leaked information is guaranteed to be below a certain threshold, and privacy amplification can negate the knowledge about the final key which that partial information imparts.

\begin{figure}
\centering
\includegraphics[width=0.9\textwidth]{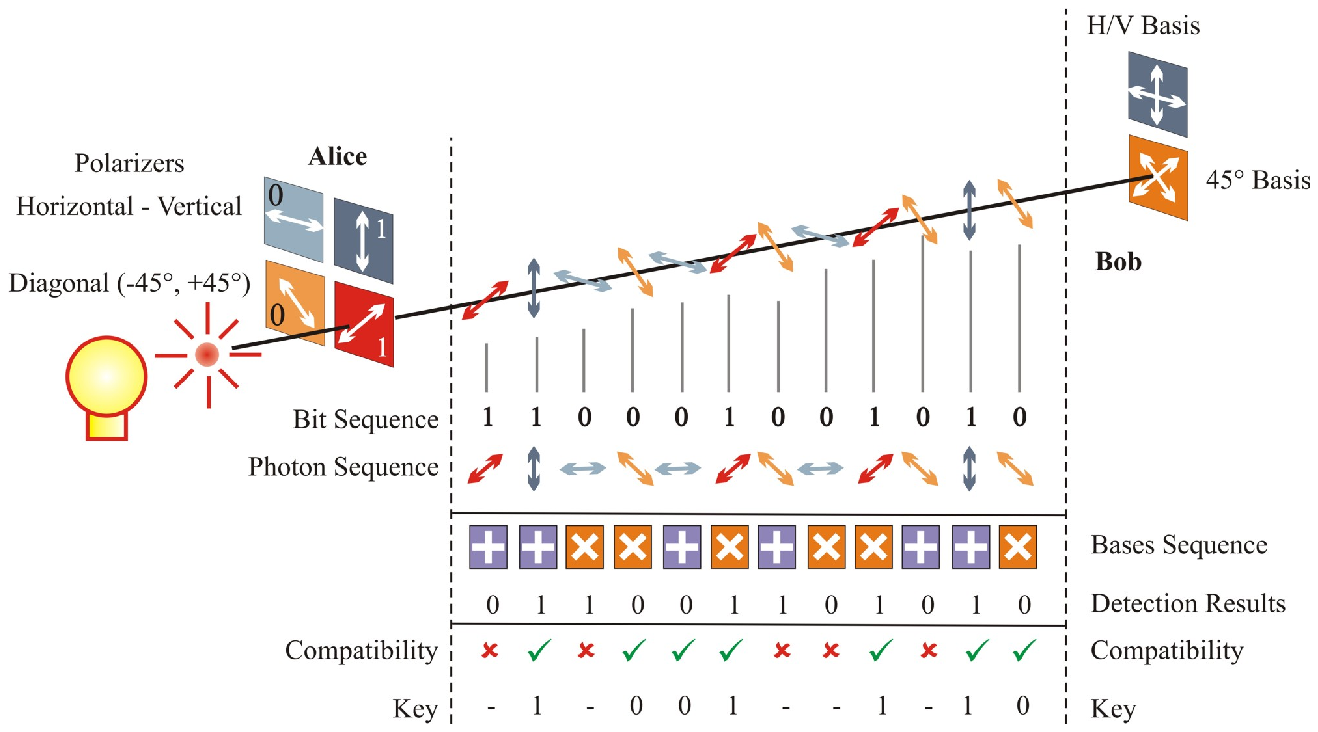}
\caption*{\small The BB84 protocol. Figure retrieved from \protect\url{http://swissquantum.idquantique.com/?Key-Sifting}.}
\end{figure}

\subsection{Modified BB84 protocols}

The best known modification of BB84 is SARG04, which adapts it for use with attenuated laser pulses \citep{SARG:04}. SARG04 is more robust than BB84 against so-called `coherent attacks', but unfortunately it performs worse against certain `incoherent attacks' \citep{Branciard:05}.

Lo, Chau, and Ardehali presents a modification of BB84 which essentially doubles its efficiency \citep{Lo:05b}. The key differences are that significantly different probabilities are assigned to the two bases so that few bits are discarded, and that key extraction is performed separately for data in each of the bases. The Cambridge--Toshiba team further improved efficiency and included decoy states, developing a new protocol called T12 \citep{Lucamarini:13}. The authors prove it to be unconditionally secure. As of February 2015, this is the protocol with which the highest ranges and secure key rates have been obtained \citep{Korzh:15}. 

Decoy state QKD comes to solve the problem that the secure key rate of a quantum key from a coherent source scales like the square of the transmittance (the proportion of photons that make it through from Alice to Bob) of the medium, and thus a secure key becomes too long to be practical when it must be transmitted for long distances. For decoy-state QKD, the key length scales like the transmittance. Three-source decoy-state QKD was what was used in \citep{Lucamarini:13}. Decoy state QKD works also with non-coherent sources such as PDC sources \citep{MaLo:08}.

Additionally, there has been work on measuring-device independent (MDI) QKD, in which Alice and Bob independently prepare phase randomized coherent pulses in one of the four BB84 states (with decoy states) and send these to an untrusted third party, Charlie. Charlie then performs Bell state measurements (BSM), and announces to Alice and Bob over a public channel the successful BSM events. Alice and Bob can get a sifted key by dropping events where they sent pulses in different bases \citep{Wang:13}. This has been implemented and gives good key rates in the laboratory \citep{Tang:14}. A further improvement has been examined, using four-source decoy states \citep{Jiang:15}.

There has been recent work to modify the BB84 protocol to deal with higher bit error rates on the sifted key, in order to distribute quantum keys for longer distances without using repeaters in a way that is compatible with optical amplification \citep{HughesNorholdt:14}.

\subsection{Continuous Variable (CV) QKD}

Continuous-variable (CV) QKD protocols employ continuous or discrete modulations of the quadratures of an electromagnetic field. CV-QKD setups rely on a coherent detection between the quantum signal and a classical reference signal, and their implementation requires only standard telecom components. They are compatible with wavelength division multiplexing, which greatly eases their deployment into telecommunication networks. They should be easier to integrate on silicon photonics chips \citep{Jouguet:13b,Kumar:14}.

\begin{sloppypar}
The bottleneck for CV-QKD is a classical cryptography problem, that of error-correction. For a long time, the range of CV-QKD was limited to 25-30km. New error-correcting codes have improved this range to 80km \citep{Jouguet:12}. SeQureNet's Cygnus module for CV-QKD features this range (\url{http://sequrenet.com/products.html}). Currently, CV-QKD keyrates are competitive with DV-QKD keyrates up to about 30km. But it may be more difficult to increase ranges for CV-QKD than for DV-QKD because security proofs for CV-QKD are penalized heavily for finite size effects.
An additional concern is that CV-QKD is a newer technology, and therefore has different vulnerabilities, some of which may be unmapped. Several potential vulnerabilities have been identified and addressed in \citep{Jouguet:13,Huang:14,Kunz-Jacques:15}.
\end{sloppypar}

Considering the above, CV-QKD should be considered a promising future technology for medium-range QKD. The state of the art for CV-QKD is surveyed in \citep{Jouguet:14}.

\subsection{Entanglement-based protocols}

There are a number of protocols involving entangled pairs of photons, chief among these being E91 \citep{E:91}.  In E91, Alice and Bob each have half of an entangled state (EPR pair, or singlet). The working concept of this scheme is that there is nothing for Eve to intercept, as the qubit state manifests only after a measurement has been made. If Eve attempts an intercept-resend attack, her measurement will break the entanglement between the photons.

The protocol proceeds as follows. Alice and Bob each choose one of two different bases to measure, with 50\% probability of choosing one basis and 50\% probability of choosing the other. After having performed their measurements, they disclose which bases they used over a public channel. If the results of measurements which were made in \textbf{different} bases violate Bell inequalities, then the state is still entangled and there has been no eavesdropping.

Despite being theoretically more secure than BB84 and its variants (fewer side-channels and thus fewer bits required for a secure key), entanglement-based protocols are not currently considered to be practical for long-range large-scale systems because of the difficulty of controlling entangled pairs caused by decoherence \citep{ScaraniKurtsiefer:14}.

\subsection{Counterfactual QKD}

Tae-Gon Noh has demonstrated that a QKD can be achieved ostensibly without sending the key through the quantum channel \citep{Noh:09}. The quantum principle in play is that the possibility of sending a photon can be detected even if the photon is seemingly not actually sent. Counterfactual QKD has been demonstrated experimentally in the laboratory \citep{Liu:11}.


\section{Real-time key extraction}

Key generation bandwidth in a pure CPU-based implementation has been shown to saturate at rates of around 1MHz \citep{Restelli:09}. High speed QKD networks routinely exceed this data rate--- for example, the NIST system generates sifted keys at around 2MHz and has a maximal capacity above 30MHz. For secure real-time practical applications, GHz data rates are anticipated. In order to shift the bottleneck from the classical computation layer back to the physical layer where it should be, hardware-based implementations have become necessary.

Sifting is computationally straightforward, and is relatively easy to perform at high speed. There is good privacy-amplification software which can work directly on a CPU-based system \citep{Zhang:14}, and the Wegman--Carter strongly universal hashing method, as used by \textit{e.g.} Id Quantique, is also good. It is the error correction step which is complicated and which sets a hard upper limit on the secure key rate.

The Cascade error-correction algorithm, developed for QKD in \citep{Brassard:94}, is the fastest at current data rates, and is best implemented in a Field Programmable Gate Array (FPGA) because it requires many simple but different logical bit-level operations. When the NIST QKD network began to exceed data rates of 1MHz, implementation of the Cascade algorithm was moved to hardware \citep{Mink:06}. The maximal throughput they were able to achieve was 12MHz in theory, but they were not able to approach that limit in a practical system due to timing jitter in their photon detectors \citep{MinkBienfang:13}. The Wuhu metropolitan area QKD network uses a similar FPGA-based system \citep{Zhang:12}.

For next-generation real-time error correction as data rates push towards the GHz mark, the Low-Density Parity-Check (LDPC) algorithm is expected to replace the Cascade algorithm for error correction \citep{Elkouss:09}. The LDPC algorithm requires 20 to 30 bytes of memory per bit of data being corrected, as opposed to 1 or 2 bytes for the Cascade algorithm. On an FPGA, LDPC implementation rates of up to 607MHz have been reported \citep{Mhaske:15}. The current fastest implementation of the LDPC algorithm runs on a GPU-based system \citep{Falcao:09} and has been tested for QKD \citep{Mateo:13, DixonSato:14}. For even faster rates, LDPC performance of 47 GHz has been reported for a custom chip implementation but not in the context of QKD \citep{Zhang:09}.

\section{Hardware: Photon sources}

A common method of encoding qubits is the use of polarized photons (less common methods include time-bin encoding \citep{Marcikic:02} and frequency encoding \citep{Zhu:10}). To preclude photon number splitting attacks, each qubit should be sent on a single photon.

The ideal single-photon source would send a single photon 100\% of the time whenever the user wishes (``on demand''), would send multiple photons 0\% of the time, and the photons it sends would be indistinguishable.

If a photon cannot be sent 100\% of the time on demand, the detector might have to be left on for a longer time, increasing `dark count' (detection of photons which were not sent to it) and thus increasing noise. If the source were to send multiple photons, then Eve would be able to intercept one photon and transmit the remaining photons to Bob, executing a \emph{photon number splitting attack}. And if photons were distinguishable, then interception one photon could give non-trivial information about another photon.

Photon sources are classified as \emph{deterministic} versus \emph{probabilistic}. A deterministic single photon source emits a single photon on demand, whereas a probabilistic source might emit more than one photon, and its photon emission timing not be entirely on demand. One should note, however, that even the most `deterministic' photon source might in practice exhibit probabilistic behaviour because, for example, photons might get lost during emission with some probability (``\emph{extraction loss}'').

A common measure for the efficiency of a single-photon source is the \emph{$2$nd order correlation function} $g^{(2)}$. If $g^{(2)}=1$, this means that the number of photons emitted by the source follows a Poisson distribution, which is the distribution one would expect from completely random and uncorrelated emissions. It is usually assumed that $g^{(2)}=1$ for attenuated lasers, although, as pointed out by the European Telecom Standards Institute, standard number GS QKD 003 Section 6.4.1 \citep{ETSI:15}, experiment hasn't always born this out and perhaps further study is necessary \textit{e.g.} when the diode is biased close to the lasing threshold \citep{Dixon:09}. The $g^{(2)}<1$ situation is referred to as \emph{photon antibunching}. In this case the probability of emitting one photon relative to the probability of emitting multiple photons is higher than in a Poisson process. The ideal state is $g^{(2)}=0$, which means that we get a single photon $100\%$ of the time.

The most common single photon sources are attenuated lasers, in which a laser beam is sent through a powerful attenuator which weakens it to the point that the probability of emitting one photon is greater than the probability of emitting multiple photons. Attenuated lasers are relatively cheap, convenient, and robust.

When higher performance (lower $g^{(2)}$) is desired, the most common single-photon sources make use of parametric down-conversion (PDC). This type of source is not on-demand, but it probabilistically produces a pair of photons, one of which can be used as a \emph{heralding photon} to instruct the detector to activate. This is a major advantage in QKD where it is important to minimize detector dark-count. The heralding photon could also be used as an entangled pair with the first photon, although here PDC makes it difficult in general to obtain the desired wavelength and phase-matching for the pair \citep{Eisaman:11}.

A promising future technology is the use of nitrogen vacancy (NV) color centers in diamond as single photon sources. An NV center is a defect in a diamond lattice which occurs when a nitrogen atom is substituted for a carbon atom, leaving a vacancy next to it. As single photon sources, NV centers are on-demand and exhibit low $g^{(2)}$. The current challenges are that they are not identical, although some tunability has been demonstrated \citep{Tamarat:06}, and that the `shelving level' reduces their efficiency. There are several proposed approaches to solving these problems (\textit{e.g.} \citep{Babinec:10}); but it is the promise of a single photon coupled with a long-lived spin qubit (the vacancy itself is an excellent qubit) that makes NV centers especially promising. Note also that must be thousands of optical defects in solids which could potentially be used for single-photon generation; only two of these have so far been seriously studied in this context \citep{Santori:10}.

\begin{figure}
\centering
\includegraphics[width=0.25\textwidth]{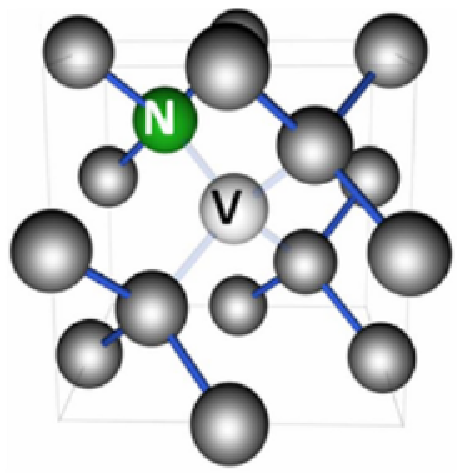}
\caption*{\small An NV center used as a single photon source. Retrieved from \protect\url{http://xqp.physik.uni-muenchen.de/research/single_photon/index.html}}
\end{figure}

There are many other single-photon sources, including single atoms, ions, and molecules, ensembles, quantum dots, nanowires, four-wave mixing, and mesoscopic quantum wells, but these do not currently seem as suitable for QKD as the sources discussed above.


\section{Transmission}

Quantum key distribution can be performed through fiber, through free space, or (experimentally) bounced off a satellite. The principles of sending photons through fiber and through free space are the same, but fiber provides a channel in which the amount of noise can be determined and even controlled to some extent, whereas the amount noise in a free-space channel is unknown (although sometimes one may try to estimate it as in \textit{e.g.} \citep{Gabay:05}) and typically is changing. Frequencies used to send photons are typically around 800nm for free space and around 1550nm for fibers. Experimental QKD usually uses dark fibers with no other signals passing through it, but real-world applications will typically involve sending messages through bright fibers which are carrying other signals. Scattering effects in bright fibers will raise the BER of Alice's transmissions, and will cause more of Alice's photons to `get lost on the way'. Despite this, by smartly time-filtering QKD photon and other communication photons, in 1992 a team from Toshiba was able to obtain a secure bit-rate of 507KHz over a 95km bright fiber, several factors of 10 over what had been achieved previously \citep{Patel:12}.

The greatest distance positive key rates have been experimentally obtained through fiber is 307km \citep{Korzh:15} and through free space is 144km between two Canary Islands \citep{Ursin:07}. The problem with free space transmission is atmospheric turbulence--- random fluctuations in the refractive index of air. One potential solution is to bounce the polarized photons off satellites. The distance to the International Space Station is 400km, but the atmospheric thickness is an order of magnitude smaller than the Canary Islands experiment. In 2014 a team from the University of Padua bounced photons off four satellites to show feasibility \citep{Vallone:14}, and China claims to have done so as well, and aims to have a dedicated QKD satellite in orbit by 2016 \citep{Yikra:14}. When such technologies become practically viable, they will significantly increase free space QKD ranges.


\section{Hardware: Photon detection}

For Bob to receive qubits from Alice in the form of single photon polarizations, Bob needs to have a good single photon detector. The main technological bottleneck in the development of practical and secure QKD systems for short to medium distances is the development of good single photon detectors. Thus, in the last few years, any improvement to single photon detection technology has immediately led to improved QKD capabilities. Our main reference for this section are \citep{Eisaman:11} and \citep{Hadfield:09}.

\begin{figure}
\centering
\includegraphics[width=0.5\textwidth]{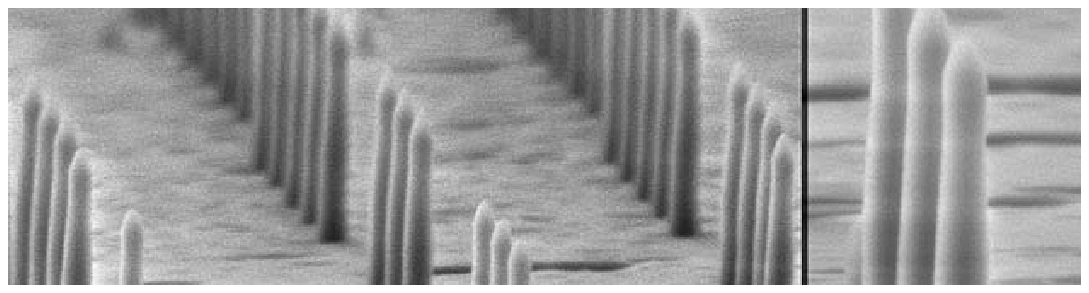}
\caption*{\small Single crystal diamond nanowires for photon detection. Retrieved from \protect\url{http://www.osa-opn.org/opn/media/Images/photocontests/gallery09_36.jpg}}
\end{figure}


An ideal single photon detector for QKD should have 100\% \emph{detection efficiency} (every photon sent to the detector should be successfully detected), 0\% \emph{dark count} (the detector should not detect photons which were not sent to it), no \emph{dead time} (the recovery time for the detector after it has detected a photon until it had detected another photon), and no \emph{timing jitter} (the time between the photon's arrival and its registration by the detector). Additionally, an ideal detector would have complete \emph{photon number resolution}, meaning that it would be able to count the number of photons it had received. It would also be \emph{asynchronous}, meaning that it need not know the arrival times of photons in advance.

Low detection capacities and high dark counts create noise in the communication channel, reducing its capacity. A low capacity channel is vulnerable to an intercept-resent attack, because it is difficult to detect eavesdropping in the presence of random noise (Section~\ref{SS:InterceptResend}). High dead time reduces the channel bit rate, and creates a vulnerability to faked-state attacks and to time-shift attacks \citep{Makarov:06, Burenkov:10,Makarov:09}. Timing jitter can lead to a leak of secret key information \citep{LamasKurtsiefer:07}. Poor or nonexistent photon number resolution creates a vulnerability to photon number splitting attacks.

A single photon detector typically works by converting a photon into a charge carrier which in turn triggers an avalanche process in a physical system which is held very close to a critical state, leading to a macroscopic current pulse.


Superconducting nanowire single photon detectors (SNSPD) are the best single photon detectors known currently. They were first developed in 2001. They have high detection efficiency (ten times better than the best semiconductor detectors), low dark count, low dead time, and low time jitter. They are also fully asynchronous. Their drawback is that they require cryogenic temperatures ($<3 K$) to operate, which makes them bulky and expensive, and has limited their uses outside the lab. There is work being done, however, on variations of SNSPD that can function at temperatures of over $20K$, making them a promising future technology for military and government applications \citep{Wang:09}.

Single photon avalanche detectors (SPADs) are the single photon detectors which are most currently used in practice. They are cheap and compact, with high detection efficiency and low time jitter. They are also fully asynchronous. The challenge in building good SPADs has been \emph{afterpulsing}, which is the phenomenon of a spontaneous dark count occurring shortly after a photon detection. If we wait until afterpulsing ends before reactivating the SPAD, then we increase the dead time. 


There has recently been dramatic progress in SPAD design. In 2013, the University of Geneva Applied Physics team developed an InGaAs negative feedback avalanche diode (NFAD) single photon detector whose performance rivals many SNSPD systems, but which operates at temperatures of approximately 150--220 K (as opposed to $<3K$ for SNSPD systems) \citep{Korzh:14}. Using these InGaAs NFADs, the same team were able in February 2015 to demonstrate provably secure QKD transmission over 307km of optical fiber, which is the current record \citep{Korzh:15}.


\section{Auxiliary systems}

\subsection{Random number generators}

A QKD system is only as good as its random number generator. If Eve can predict Alice and Bob's random choices, they she can read the entire key. The entire selling point of Quintessence Technologies is their random number generators.

CPU-based random number generators are trusted for many classical cryptography tasks, and are implemented in most operating systems. The numbers they produce are not truly random, however, and therefore they are usually referred to as \emph{pseudorandom number generators}. When higher speeds are required and when stronger random numbers are needed, hardware-based implementations are preferred. These come in two flavours--- they either use filtered random physical processes within the FPGA as random number seeds \citep{Tsoi:07, Kwok:06}, or they use less random seeds and strong permutations \citep{Alimohammad:08,Cheung:07,Xiang:09}. Currently, both alternatives are considered cryptographically equivalent.

In QKD, in order to physically guarantee unconditional security, quantum effects are desired for use as true random number generators (TRNG). A quick and dirty way to do this, for Alice at least, is to send an unpolarized single photon through a beam splitter--- if it comes out one end then count that as a zero, and if it comes out of the other end count it as a 1. A more sophisticated version of this scheme which eliminates this bias is marketed by Id Quantique, and another by Quintessence Technologies, and reaches rates of 16 KHz. Looking into the future, an experimental idea with great promise is to use quantum vacuum fluctuations for high bandwidth truly random number generation of up to 100GHz \citep{Jofre:11}.

We note that TRNGs arise as commercial spinoffs of QKD projects.

\subsection{Memories and repeaters}

\begin{figure}
\centering
\includegraphics[width=0.8\textwidth]{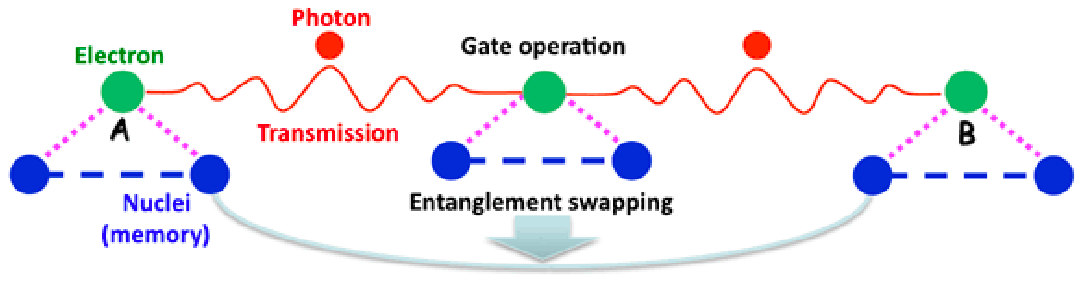}
\caption*{\small The concept of a quantum repeater. Retrieved from \protect\url{http://www.uqcc.org/research/index.html}}
\end{figure}

To extend the range of QKD beyond a few hundred kilometers, quantum repeaters will be necessary, which in turn will require quantum memories. A quantum memory absorbs a photon, stores its quantum states for as long as possible, and releases it on demand. A key feature is that it does not break entanglement. The primary candidates for practical quantum memories for QKD in the near future are Raman gas based quantum memories \citep{Simon:10} and quantum memories using NV centers \citep{deRiedmatten:15}. The advantages of the former include its greater capacity, while the advantages of the latter include that it is solid-state and that it allows longer storage times. It is still unclear which of these approaches will be best.

It is still unclear which repeater technology will be best, although the first quantum repeaters which outperform direct transmission will probably be based on atomic ensembles, linear optics, and photon counting \citep{Sangouard:11}.

Quantum memories and repeaters are expected to become a commercial technology within 10-15 years.

\section{Attacks}

While the protocols of QKD operating under certain conditions are unconditionally secure, practical implementations have been successfully attacked. While none of these attacks is fatal to the QKD concept--- effective countermeasures to each attack have been devised--- it is generally agreed that the security of each setup should be the object of a dedicated study whose goal is to find and patch up all vulnerabilities \citep{ScaraniKurtsiefer:14}.

\subsection{Intercept-resend}\label{SS:InterceptResend}

The simplest and most direct attack against BB84 and its relatives is for Eve to intercept a photon sent by Alice to Bob, to measure that photon, to prepare her own photon encoding the bit which she measured, and to send that photon off to Bob. Because Eve doesn't know in which basis Alice's photon was sent, she'll measure in the wrong basis approximately half of the time and she will send a photon in the wrong basis to Bob approximately half of the time. An intercept-resend attack thus introduces a bit error rate (BER) of around 25\%, although weaknesses in certain practical systems allow modified versions of this attack to introduce BERs of 19.7\% \citep{Xi:10}. Because acceptably BERs in commercial systems are around 8\%, practical QKD is indeed secure against pure intercept-resent attacks, which are caught during the error-correction key-establishment phase--- if the error rate is too high then Eve has been there.

\subsection{Photon Number Splitting (PNS)}\label{SS:PNS}

Due to hardware limitations, most photon sources used in QKD are not true single photon sources in the sense that there is a non-negligible probability that they will generate multiple photons to transmit a single qubit. If Alice sends two or more identical photons to Bob, then Eve can split off one photon and send the remaining photons through. Eve stores the qubit she has learnt in quantum memory until Alice has revealed her encoding basis. Then Eve measures her photons in the correct basis and gains information about the key.

A successful photon number splitting attack requires sophisticated equipment--- Eve must be able to count photons and to split off just one to quantum memory while sending others through. Moreover, various countermeasures have been developed. Better single-photon sources and modifications of the BB84 protocol, such as for instance SARG04, make successful PNS attacks much more difficult to carry out. Another solution is to use decoy states, in which photons are randomly sent at different intensities. The security of decoy state QKD against PNS attacks was proven in \citep{Lo:05}. Because a successful photon number splitting attack is much more difficult to carry out against decoy-state QKD, we can use attenuated lasers as photon sources when transmitting keys, increasing secure key-rates \citep{Yuan:07}. The current state-of-the-art for decoy-state QKD is 320MHz over a 200km fiber, yielding a 15Hz secure key rate \citep{Liu:10}.


\subsection{Timing attacks}\label{SS:TimingAttack}

When different light sources are used for beams in different polarizations, and/or different detectors are used to make different measurements, it may be possible to `listen in' to which bit was sent or to which bases was used without actually intercepting a photon. Such side-channels were evident already in the first implementations QKD, as noted by Brassard \citep{Brassard:05}:

\begin{quote}
The funny thing is that, while our theory had been serious, our prototype was mostly a joke. Indeed, the
largest piece in the prototype was the power supply needed to feed in the order of
one thousand volts to Pockels cells, used to turn photon polarization. But power
supplies make noise, and not the same noise for the different voltages needed for
different polarizations. So, we could literally hear the photons as they flew, and
zeroes and ones made different noises. Thus, our prototype was unconditionally
secure against any eavesdropper who happened to be deaf ! :-)
\end{quote}

It is therefore critical to the security of the QKD system that different light-sources and detectors be indistinguishable to Eve. One particular vulnerability is that different light sources and detectors may not be perfectly synchronized, so that Eve can figure out which detector clicked, for example, by examining the time signature publicly announced by Bob in order to determine which photon he detected of the photons sent by Alice. Such an attack could read-off $\geq 25\%$ of the key for a detector mismatch of $0.5$ nanoseconds, an amount that could easily go unnoticed \citep{LamasKurtsiefer:07}. An attempt to carry out such an attack against a commercial system was unsuccessful because of several practical difficulties \citep{Zhao:08}.

\subsection{Trojan attacks}\label{SS:Trojan}

\begin{figure}
\centering
\includegraphics[width=0.5\textwidth]{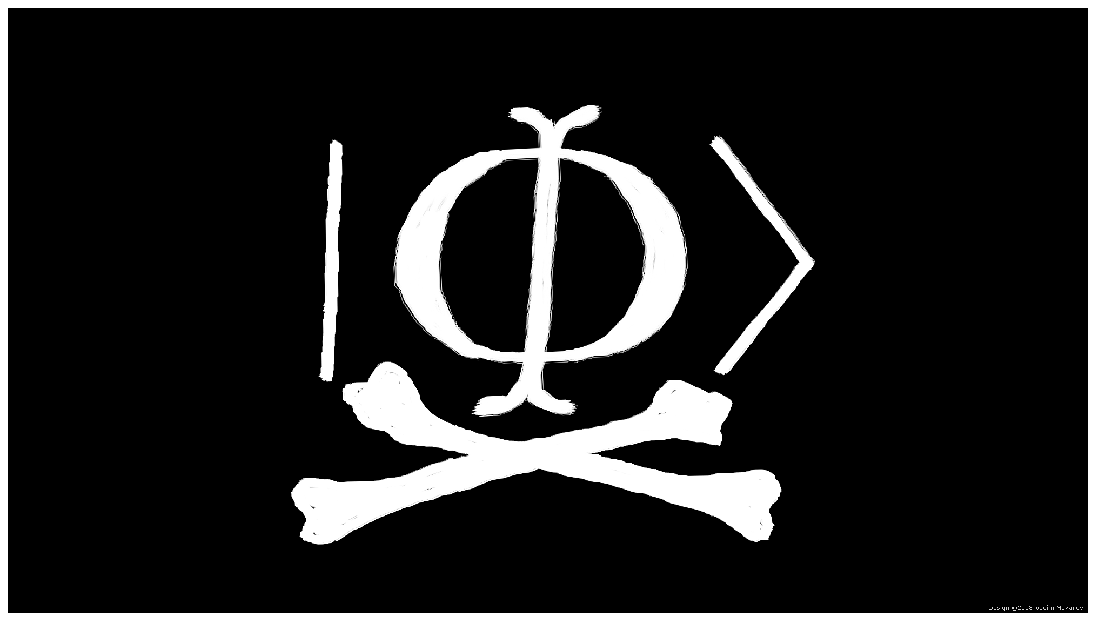}
\caption*{\small The logo of the Quantum Hacking Lab. Retrieved from \protect\url{http://www.vad1.com/lab/}}
\end{figure}

In a Trojan attack, Eve shines bright light at either Alice or Bob, determining which base was used by analyzing the reflection. The Trojan attack has successfully read off complete keys both in the lab \citep{Gerhardt:11} and also of commercial QKD systems, QPN-5505 from MagiQ Technologies and Clavis2 of IP Quantique \citep{Lydersen:10}. This has been the most powerful and the best-performing hack on QKD so far. Although these specific attacks can be protected against, Trojan attacks using pulses of different wavelengths may still be able to hack complete keys, and we still do not know the full scope of the vulnerability of practical QKD systems to Trojan attacks \citep{Jain:14,Jain:15}.

\subsection{Other side-channel attacks}

Many other attacks against QKD systems have been investigated, and new vulnerabilities are periodically discovered. Some of these attacks (denial of service, man-in-the-middle,\ldots) can be carried out against classical systems as well, and the vulnerability of QKD to these attacks is identical to the vulnerability of any classical protocol. Other attacks which take advantage of a weakness in an auxiliary system--- \textit{e.g.} a randomization attacks, in which the random bases are successfully computed by Eve because the random number generator is faulty--- can be counteracted by using better hardware. Of course each side-channel attack must be investigated and ruled out for each setup.

\section{Conclusion}

Quantum Key Distribution (QKD) is a modern form of trusted courier, which in principle allows Alice to communicate a message to Bob with complete confidence that the message will not be eavesdropped on during transmission. Real-life QKD security, however, is a quantitative issue, and each setup should be individually studied to ensure its security. QKD is currently a focus of interest for many private, governmental, and military groups all over the world.

Current state of the art setups still use the first QKD protocol, BB84, and its variants. The Cascade protocol is still the fastest for error correction, but LDPC is expected to overtake it as key rates rise. In both cases, hardware implementation using FPGA's is the current state of the art and is likely to remain so for the next decade at least.

Qubits are typically transmitted as polarized photons. Decoy-state QKD using attenuated laser pulses are the current state of the art photon sources, despite not being true single photon sources. NV centers are a promising future technology. Polarized photons can be transmitted through fiber, through air (free space), or bounced off satellites. There are various attempts to send polarized photons via bright fibers through which other messages are travelling, but the key rates being obtained are still quite low.

Photon detectors are the main technological bottleneck for practical QKD. The current state of the art are InGaAs NFAD's. A promising future technology are SNSPD's, which currently require cryogenic temperatures to operate, but future SNSPD's be able to operate at above 20K.

Memories and repeaters, which are thought to be required for QKD at ranges over around 400km, are still in the experimental stage, and it is too early to say which technology will be best.

Security of QKD is a well-studied field, and there have been numerous attempts to attack QKD implementations both using standard attacks and also using side-channel attacks. Only one of these attacks, a Trojan attack, has successfully stolen a secret key, and the vulnerability it highlights can be plugged. QKD of course has the same vulnerability to classical attacks such as denial-of-service and man-in-the-middle as classical implementations.

QKD is an exciting emerging technology which is beginning to enter the marketplace. We expect it to be successful in its own right, and also to serve as a stepping stone towards greater and higher goals in quantum communication and computation.

\section{Acknowledgement}

The author wishes to thank Judith Kupferman for useful comments, and wishes to thank Paul Jouguet at Takehisa Iwakoshi for corrections and references.

\bibliographystyle{rspublicnat}

\end{document}